\documentclass[border=2mm]{article}
\usepackage{comment}
\usepackage{float}
\usepackage{graphicx}
\usepackage{hyperref}
\newcommand*{\doi}[1]{\href{http://dx.doi.org/#1}{doi: #1}}
\usepackage{geometry}

\hypersetup{
  colorlinks   = true, %Colours links instead of ugly boxes
  urlcolor     = blue, %Colour for external hyperlinks
  linkcolor    = blue, %Colour of internal links
  citecolor   = red %Colour of citations
}

\usepackage[bitstream-charter]{mathdesign}
\DeclareSymbolFont{usualmathcal}{OMS}{cmsy}{m}{n}
\DeclareSymbolFontAlphabet{\mathcal}{usualmathcal}

\begin{document}

\begin{center}{\Large \textbf{
Charmonium and bottomonium spectroscopy at Belle II\\
}}\end{center}

% TODO: write the author list here. Use initials + surname format.
% Separate subsequent authors by a comma, omit comma at the end of the list.
% Mark the corresponding author with a superscript *.
\begin{center}
  A. Thampi\textsuperscript{1*}
  on behalf of the Belle II Collaboration
%Aah B. Cee\textsuperscript{2} and
%Gee K. See\textsuperscript{3$\star$}
\end{center}

\begin{center}
{\bf 1} Forschungszentrum Juelich, Juelich, Germany
%\\
%{\bf 2} Affiliation2
%\\
%{\bf 3} Affiliation2
\\
*a.thampi@fz-juelich.de
\end{center}

\begin{center}
\today
\end{center}

% For convenience during refereeing (optional),
% you can turn on line numbers by uncommenting the next line:
%\linenumbers

%\definecolor{palegray}{gray}{0.95}
%\begin{center}
%\colorbox{palegray}{
%  \begin{tabular}{rr}
%  \begin{minipage}{0.1\textwidth}
%    \includegraphics[width=22mm]{Logo-DIS2021.png}
%  \end{minipage}
%  &
%  \begin{minipage}{0.75\textwidth}
%    \begin{center}
%    {\it Proceedings for the XXVIII International Workshop\\ on Deep-Inelastic Scattering and
%Related Subjects,}\\
%    {\it Stony Brook University, New York, USA, 12-16 April 2021} \\
    %\doi{10.21468/SciPostPhysProc.?}\\
%    \end{center}
%  \end{minipage}
%\end{tabular}
%}
%\end{center}

\section*{Abstract}
The Belle II experiment at the SuperKEKB energy-asymmetric $e^{+}e^{-}$ collider is an upgrade of the $B$ factory facility at KEK in Tsukuba, Japan. The experiment began operation in 2019  and aims to record a factor of 50 times more data than its predecessor. Belle II is uniquely capable of studying the so-called "XYZ" particles: heavy exotic hadrons consisting of more
than three quarks. First discovered by Belle, these now number in the dozens, and represent the emergence of a new category within quantum chromodynamics. We present recent results
obtained from Belle II data, and the future prospects to explore both exotic and conventional quarkonium physics.
  

%\vspace{10pt}
%\noindent\rule{\textwidth}{1pt}
%\tableofcontents\thispagestyle{fancy}
%\noindent\rule{\textwidth}{1pt}
%\vspace{10pt}

\section{Introduction}
\label{sec:intro}
Hadrons are composite particles, made up of two or more quarks, or a combination of quarks and gluons, or merely gluons. They can be classified in different categories, depending on the constituent elementary particles inside them. Mesons ($q\bar{q}$) and baryons ($qqq$) are the conventional hadrons according to the Gell-Mann Zweig idea of the constituent quark model.
However we know that more exotic combinations exist. Models motivated by quantum chromodynamics (QCD) predict  the existence of hadrons with more complex structures than simple  mesons or baryons.

The charmonium and bottomonium spectrum were well established from experiments, and in a good agreement with predictions from potential models. In 2003, the $X(3872)$ \cite{ref:x3872belle}
was first observed by the Belle experiment. Then several new resonances were observed from the $B$ factories, LHC experiments, BESIII  and others, for which more models were elaborated.
But so far the information has been insufficient to frame all these new states in a unique model.

\section{SuperKEK and Belle II}
\begin{figure}[h]
  \centering
  \includegraphics[width=0.4\textwidth]{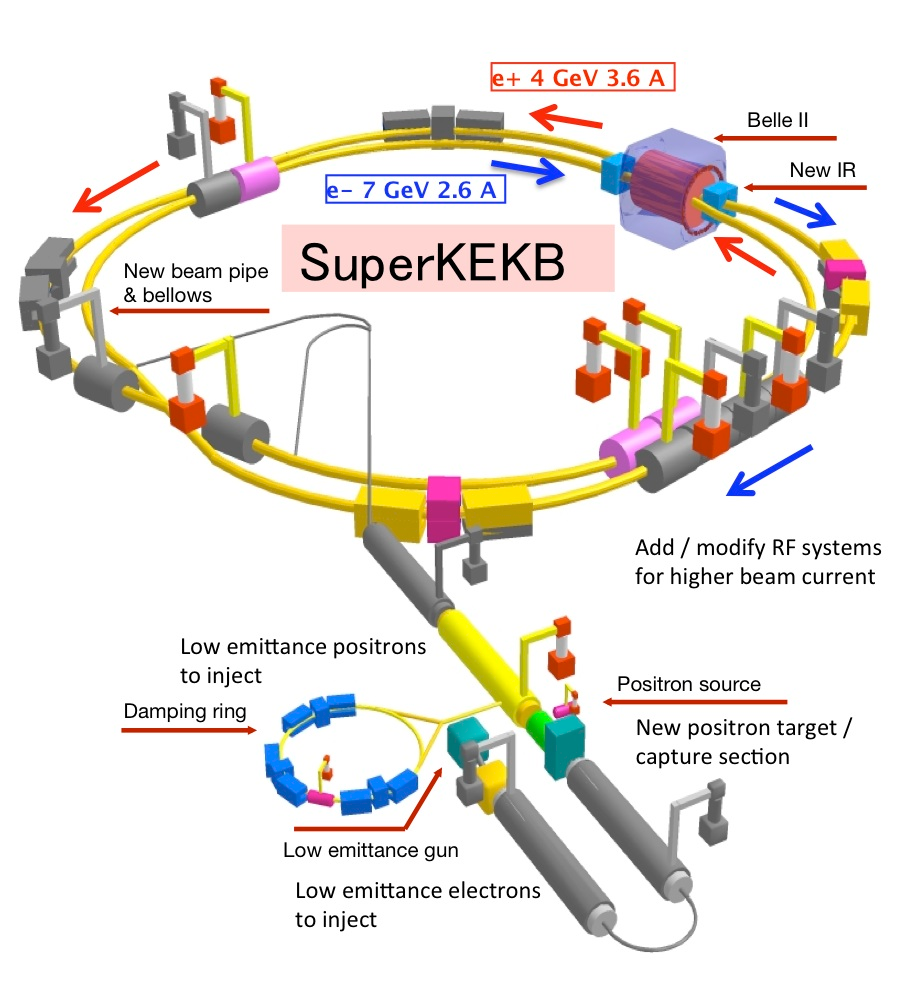}
  \includegraphics[width=0.45\textwidth]{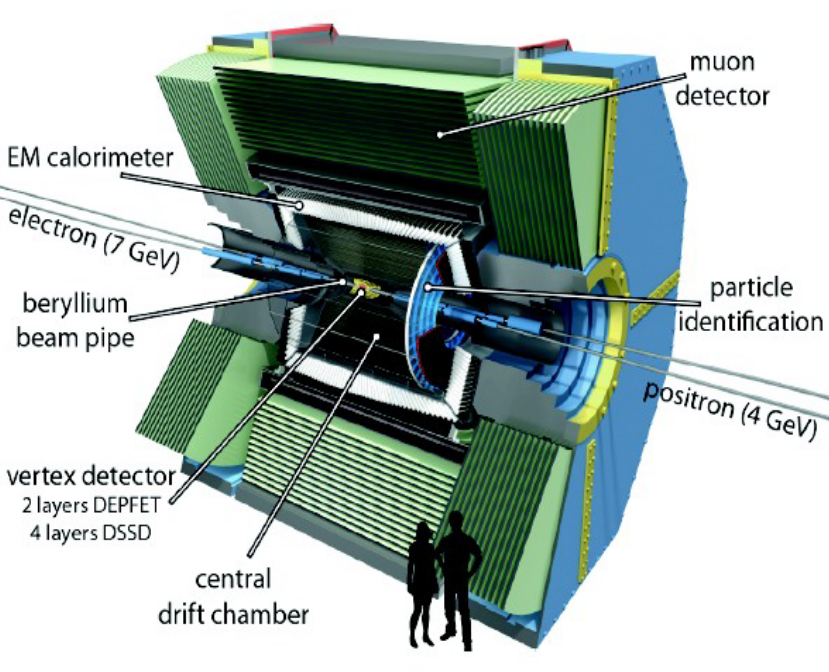}
  \caption{Schematic representation of the SuperKEKB rings (left) and the Belle II detector (right).}
  \label{fig:kek}
\end{figure}

In order to constrain theoretical models and understand the new charmonium and bottomonium states, we need data. LHCb has so far performed very well, but is limited in analyzing radiative decays involving low-energy photons.  
SuperKEKB \cite{ref:superkekb} is an upgrade of the already existing KEKB facility in Tsukuba, Japan, where an asymmetric $e^{+}e^{-}$ collider running at the center of mass (CM) energy of the  $\Upsilon(4S)$ resonance ($\sqrt{s}= 10.58$ GeV) is designed to collect up to 50 ab$^{-1}$ of integrated luminosity data. The Belle II \cite{ref:belleii} detector is placed around the interaction point, and so far has accumulated approximately 120 fb$^{-1}$ of data. An illustration of the SuperKEKB and the Belle II detector is shown in figure \ref{fig:kek}.   

\section{Studies of charmonium at Belle II}

\subsection{Search for the $X(3872)$}
\begin{figure}[H]
  \centering
  \includegraphics[width=0.95\textwidth]{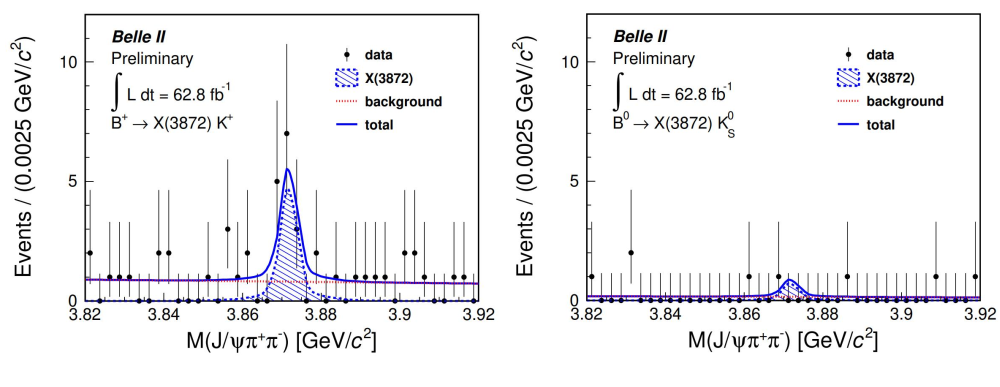}
  \caption{Distribution of $J/\psi \pi^{+} \pi^{-}$ invariant mass with a fit to the $X(3872)$ signal superimposed for the charged (left) and the neutral (right) $B$ channels.}
  \label{fig:x3872}
\end{figure}
The $B^{\pm/0} \to J/\psi \pi^{+} \pi^{-} K^{\pm/0}$ decay mode is studied to search for the $X(3872)$ state. Using the available $62.8$ fb$^{-1}$ of data, Belle II re-discovered the  $X(3872)$, finding in the charged $B$ channel  $14.4 \pm 4.6$ events \cite{ref:belleiiphybook}. The results are shown in figure \ref{fig:x3872}. The most promising channel to analyze the $X(3872)$ is $B^{\pm/0} \to D^0 \bar D^0 \pi^0 K^{\pm/0}$; however, statistics are not yet sufficient for such analysis at Belle II. The main goal of this analysis will be the measurement of the total width of the $X(3872)$.
Belle II might be able to distinguish among different parametrizations, $i.e.$ the Flatt\'e or Breit Wigner parameterization.  Monte Carlo simulations performed of the measurement
of the $X(3872)$ width in the $D^{0} D^{0} \pi^{0}$ channel demonstrate that Belle II can measure an upper limit down 200 keV, once the full integrated luminosity of
50 ab$^{-1}$ is available.

\subsection{Study of $c\bar{c}$ processes in ISR}
\begin{figure}[H]
  \centering
  \includegraphics[width=0.48\textwidth]{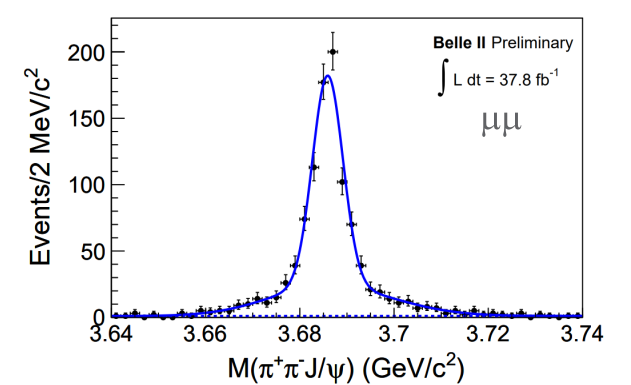}
  \includegraphics[width=0.48\textwidth]{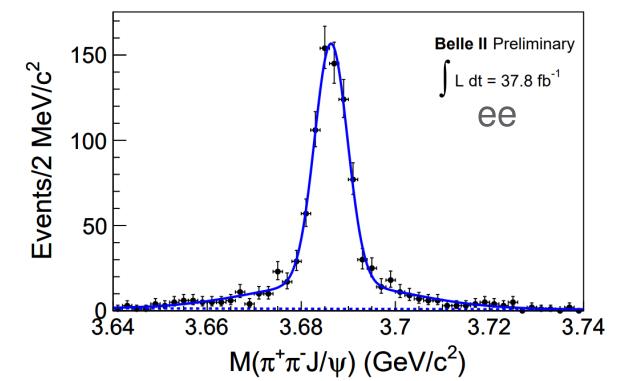}
  \caption{$\psi(2S)$ state peaking in the $J/\psi \pi^{+} \pi^{-}$ invariant mass distribution for the ISR process  $e^+ e^- \rightarrow \pi^{+} \pi^{-} J/\psi$.}
  \label{fig:ccisr}
\end{figure}

Initial state radiation (ISR) processes are a unique physics opportunity at $B$ factories, providing a clear signature of a resonant state - if it is observed - since the quantum number of such a state must be the same of the ISR photon, $i.e.$ $J^{P~C}$ = 1$^{-~-}$. A clear observation of $\psi(2S)$ states is possible at Belle II in the ISR process  $e^+ e^- \rightarrow \pi^{+} \pi^{-} J/\psi$. The $J/\psi$ meson is reconstructed from $e^{+}e^{-}$ or $\mu^{+} \mu^{-}$ within a mass range of 75 MeV$/c^2$ from the nominal value \cite{ref:pdgjpsi}.
A selection on the missing mass squared ($|MM^{2}(J/\psi \pi^{+} \pi^{-})| < 2$ GeV$/c^2$) is required. The results acquired by using data corresponding to an integrated luminosity of  37.8 fb$^{-1}$ are shown in figure \ref{fig:ccisr}.

\section{Studies of bottomonium at Belle II}
\begin{figure}[H]
  \centering
  \includegraphics[width=0.7\textwidth]{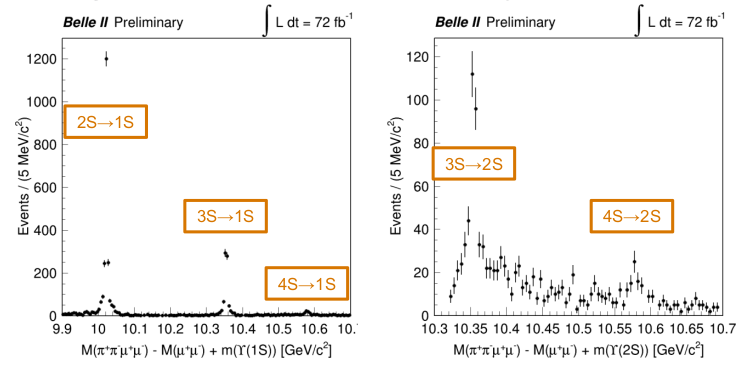}
  \includegraphics[width=0.35\textwidth]{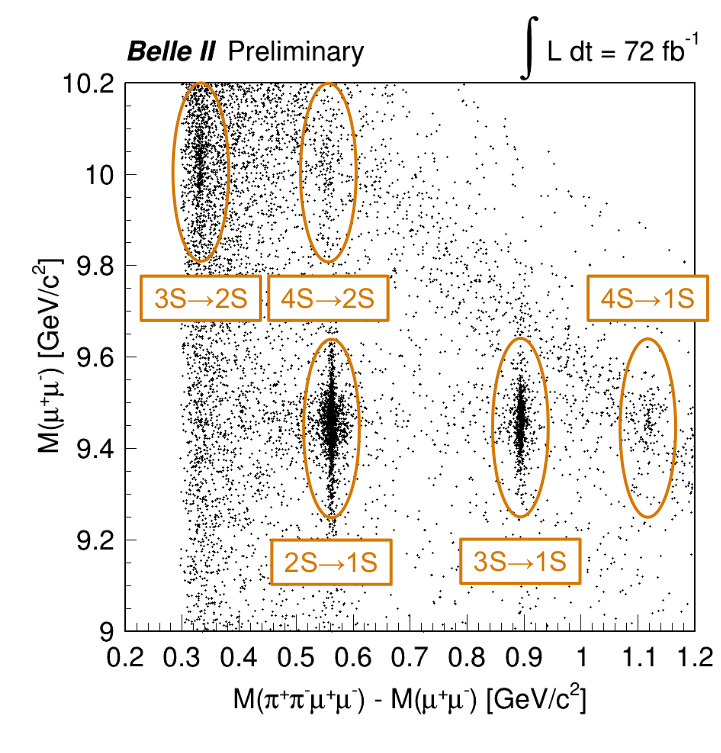}
  \caption{Results for the ISR bottomonium search using 72 fb$^{-1}$ of data.}
  \label{fig:bbisr}
\end{figure}

In the $e^{+}e^{-} \gamma_{ISR} \to \pi^{+} \pi^{-}$ ISR production, the transitions $\gamma_{ISR} \Upsilon(2S) \to \pi^{+} \pi^{-} \Upsilon(1S)(\ell^{+}\ell{^-})$ and
$\gamma_{ISR} \Upsilon(3S)$ $ \to \pi^{+} \pi^{-} \Upsilon(1S,2S)(\ell^{+}\ell^{-})$ are observed. Also the direct transitions
$\Upsilon(2S) \to$ $\pi^{+} \pi^{-}\Upsilon(1S)(\ell^{+}\ell^{-})$ are observed. The acquired results using 72 fb$^{-1}$ of data collected by Belle II are shown in figure \ref{fig:bbisr}.
This preliminary study, conducted on a very small sample compared to the whole planned data sets, already shows the potential in this field.
A plan to study such $\Upsilon(nS) \to \Upsilon(mS)$ transitions in detail has been approved and the goal is the study of both conventional and exotic states, as well as lepton-flavor violation as probe of new physics.

\section{Conclusion}
The Belle II experiment is performing well. In the first two years of data taking we have already collected half of the BaBar integrated luminosity, which was collected  over eight years.
With a 50 ab$^{-1}$ data sample, interesting analyses are expected in the charmonium and
bottomonium sectors. So far we can show only re-discovery channels, testifying to the good performance of the detector.  

%You must include a conclusion.

%\section*{Acknowledgements}
%Acknowledgements should follow immediately after the conclusion.

% TODO: include author contributions
%\paragraph{Author contributions}
%This is optional. If desired, contributions should be succinctly described in a single short paragraph, using author initials.

% TODO: include funding information
%\paragraph{Funding information}
%Authors are required to provide funding information, including relevant agencies and grant numbers with linked author's initials. Correctly-provided data will be linked to funders listed in the \href{https://www.crossref.org/services/funder-registry/}{\sf Fundref registry}.

% \bibliographystyle{SciPost_bibstyle} % Include this style file here only if you are not using our template
%\bibliography{SciPost_Example_BiBTeX_File.bib}

%\nolinenumbers

\end{document}